# Fundamental Radar Properties: Hidden Variables in Spacetime

Andrew K. Gabriel

*Jet Propulsion Laboratory*
*California Institute of Technology*
*Pasadena, California 91109*

## Abstract

A derivation of the properties of pulsed radiative imaging systems is presented with examples drawn from conventional, synthetic aperture, and interferometric radar. A geometric construction of the space and time components of a radar observation yields a simple underlying structural equivalence between many of the properties of radar, including resolution, range ambiguity, azimuth aliasing, signal strength, speckle, layover, Doppler shifts, obliquity and slant range resolution, finite antenna size, atmospheric delays, and beam and pulse limited configurations. The same simple structure is shown to account for many interferometric properties of radar - height resolution, image decorrelation, surface velocity detection, and surface deformation measurement. What emerges is a simple, unified description of the complex phenomena of radar observations.

The formulation comes from fundamental physical concepts in relativistic field theory, of which the essential elements are presented. In the terminology of physics, radar properties are projections of hidden variables - curved worldlines from a broken symmetry in Minkowski spacetime - onto a time-serial receiver.

# Introduction and Literature Review

The very rich and complex phenomena associated with radar returns are used in a corresponding multitude of applications. Earthquake deformations, missile detection, weather prediction, wind measurement, topographic mapping, ocean currents, target tracking, submarine detection, aircraft guidance, air control, and planetary mapping are only a partial list. Many of these radar properties appear to be unrelated - for example, resolution and ambiguity, and are usually treated as separate topics. In this paper, a simple spacetime formulation is applied to multiple types of radar observations. Resolution is seen to be a limiting case of range ambiguity; other radar phenomena, including speckle, obliquity and slant range corrections, interferometric properties, azimuth ambiguity, signal strength, atmospheric delays, finite antenna corrections, beam and pulse limited configurations, and Doppler shifts emerge as different facets of the same simple underlying geometric structure, which is shown to be the curvature of the relativistic worldlines associated with the radiation, curvature that itself originates in a simple broken symmetry. The very complex properties of radar returns result from the projection of those relatively simple higher dimensional ("hidden") variables onto the lower dimensional time-serial receiver. In conventional terminology, this projection is much the same idea that the projection (shadow) of the edges of randomly oriented three-dimensional cube onto a two-dimensional plane can produce complex shapes even though the higher dimensional cube is very simple.

A full discussion of radar, including the SAR concepts, may be found in (1). The usual treatment of temporal radar returns and ambiguity involves the Woodward ambiguity function ((2) and citations), but this approach deals with coded waveforms, not the geometric properties considered here. Extensive radar calculations that involve some combination of space and time variables in the orthodoxy of signal processing may be found in, for example (3). One radar paper was found using geometric methods (4), but is a derivation of Doppler related waveform design. While time-reversal techniques (below) appear to be common in seismic sounding research, searches of various combinations of "time", "reversed", and "radar" produced almost no relevant returns on the ISI Web of Science citation search system. Searches for "radar and world line" came up empty, as did searches for "radar and hidden variables"; searches for "radar and relativity" produced a scant few references (e.g. (5)), all of which dealt with radar as a luminal source in special or general relativity, rather than the imaging properties of concern here.

# Light Cones and World Lines in Radar

In the theory of special relativity (5), Minkowski spacetime is a fundamental and simple concept. Space becomes a four dimensional entity where the extra dimension is time. Scaling this dimension by c, the constant speed of light, yields the "timelike" dimension which is measured in units of ct. Any stationary point in Euclidean space exists for all time, and thus in spacetime becomes a line, or "worldline" that runs parallel to the ct axis. In the simplest case of a two dimensional (x and ct) spacetime, a object moving with constant speed v in the x ("spacelike") direction has a worldline that is straight but pitched at an angle $\theta$ to the ct axis where $\tan \theta$ = v/c. Similarly, an accelerating object has a curved worldline (for example, a parabola for a constant spacelike acceleration).

The worldlines of radiation have unique properties. Since all radiation moves in vacuum at constant speed c, a light ray in two dimensional spacetime has straight worldlines at the two possible angles with the ct axis where tan $\theta$ =1 ($\theta = \pm\pi/4$). In three dimensional spacetime (x,y,ct), light rays originating at some point x,y can be anywhere on a cone with its vertex at (x,y), its axis of symmetry parallel to the ct axis, and its central angle equal to 2($\pi/4$) = $\pi/2$. An isotropic radiator emits a spherical wave in Euclidean (x,y,z) space which appears in three dimensional (x,y,ct) spacetime as light rays originating at the vertex and propagating uniformly along the cone, usually called a "light cone". Fig.1 shows a light cone intersecting the (x,t) plane. Similarly, incoming light incident on some point (x,y,z) in space must also be on a light cone with its vertex at that point, but the cone is rotated by $\pi$ so the open end faces the -ct direction (the other solution of tan $\theta$ = 1 where $\theta = \mp\pi/4$). The symmetry of the construction suggests the notion that reception can be viewed as time-reversed transmission; similarly, transmission can be viewed as time-reversed reception. Since nothing can exceed the speed of light, a light cone divides spacetime into regions that cannot communicate without violating causality; for this reason it is sometimes called the "causal cone".

Radiation always propagates on a radiative worldline; however, in general relativity, spacetime is not rectilinear but can become curved, and the worldlines of radiation also become curved. For example, the presence of mass changes the local curvature of spacetime, so nearby radiative worldlines are not straight. In conventional terminology, light is deflected by the presence of mass; in the most extreme scenario, a black hole so severely warps spacetime that light can become trapped. Much of the interesting phenomena in relativistic astrophysics occurs as a result of curving worldlines, and it is shown herein that this is also true of much of radar phenomena.

## Spacetime Components of a Radar Observation

A simple radar antenna at some height $z_0$ emitting a spherical pulse of radiation of duration $\tau_0$ produces an illumination pattern over flat (planar) ground that begins as a single nadir point (x≡0, y≡0) at time $t_0 = z_0/c$, expands radially in the (x,y) plane into a circle until time $t_0 + \tau_0$; it then becomes a radially propagating annulus with an asymptotic spatial radial width of $c\tau_0$. The dimensionless functional form of this illuminated area is denoted $a_T(r,t)$ where r is the radial distance in the x,y plane from the nadir point. Heuristically, the transmitter creates an expanding ring-shaped illuminated area or "window" $a_T(r,t)$ on the ground. Specifically, for the present purposes, $a_T(r,t)$ is a function that has a value of unity (except when it is a $\delta$-function) when there is illumination present at some (r,t) and zero otherwise. In two dimensional spacetime, this window is the intersection of the light cone originating at the radar with the (x,t) plane; this intersection is the hyperbola $a_T(x,t)$ (note the single spatial dimension x instead of r; extension to higher dimensions is straightforward). The light cone and hyperbola are depicted in Fig1.

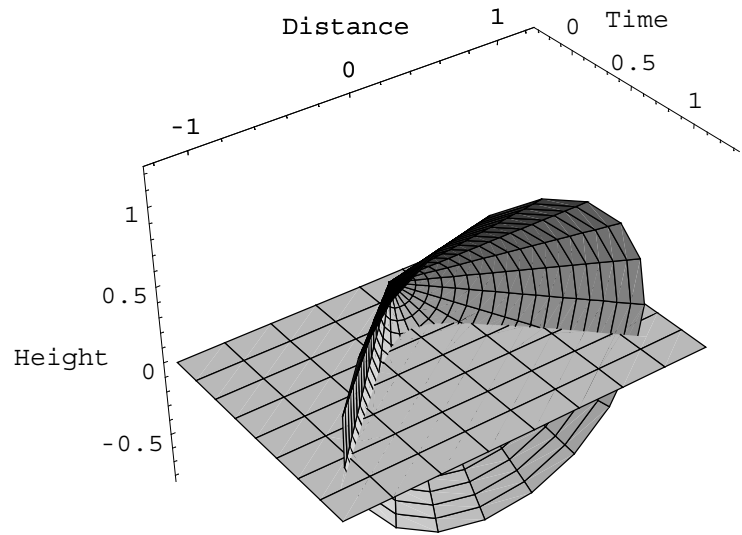

**Figure 1.** A light cone in Minkowski space originating at some arbitrary height $z_0$ above the x,t plane forms a hyperbola where it intersects the plane. Note the plane in the figure contains only *one* scene spatial dimension, the distance x.

An incoming light cone $a_R$ (x,t) looks identical to Fig.1 except that it is rotated by $\pi$ around its own vertex in the direction parallel to the (x, ct) plane . The two hyperbolas of intersection can be identified as the worldlines of the transmitter and receiver "windows" *as they appear to move in the x direction on the ground*. A shift of one cone toward the other along the time direction (by an amount called $\tau$ below) will result in a different circle of intersection in the x,z (distance, height) plane, and the hyperbolas in the x,t plane will also intersect if $\tau$ exceeds 2 $z_0$/c, as shown in Fig.3. The locus of points where the two cones intersect represent the areas of spacetime common to both $a_T$ and $a_R$ where causality allows communication to occur between the transmitter and receiver. This is shown in Fig.2.

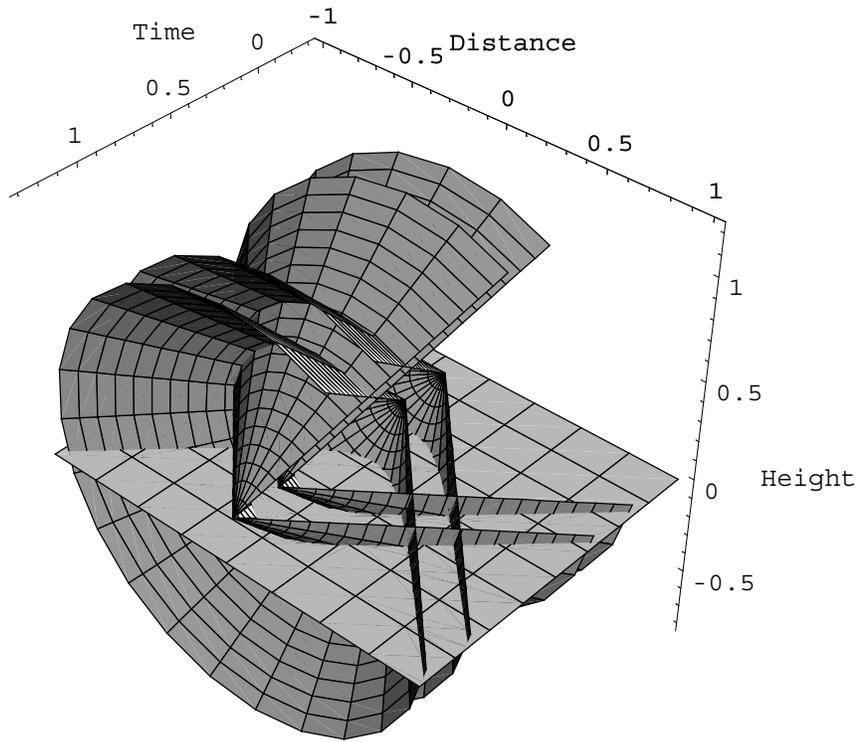

**Figure 2.** Intersecting light cones of the transmitter and receiver. The cone vertices, at height $z_0$ above the (x,t) plane are separated in time by $\tau_0$, the pulse length, for like-looking cone pairs and by $\tau$, the convolution offset, for opposite-looking pairs measured from the outermost vertices. The hyperbolas in the (x,ct) or (distance, time) plane in the figure also intersect, forming the image resolution element (rezel) shown as the diamond shape.

Looking in the x,t plane at the intersecting hyperbolas, the simplest one-dimensional case of a very narrow radiated pulse is $a_T = \delta(x(t))$, where

$$x(t) = \sqrt{((ct)^2 - z_0^2)} \qquad \text{Eq.1}$$

X(t), the domain where $a_T(x,t)$ and $a_R(x,t)$ are nonzero is depicted in Fig.3; the curvature of these worldlines is caused by the square root in Eq.1.

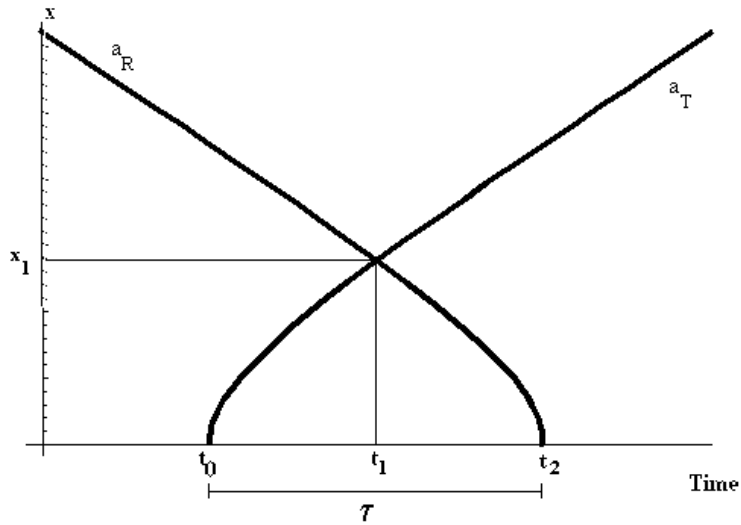

**Figure 3**. Along-ground distance of incoming and outgoing radar "window" world lines $a_T$ and $a_R$ for a $\delta$-function pulse.

.

In the (x,ct) plane, the world line of an outgoing or incoming pulse is, in the asymptotic far range, a straight diagonal line (at anle $\pi/4$ with the ct axis) where the radiation is, in effect, propagating parallel to the ground. The world line of a point target at $x_1$ would be a horizontal line; the world line of the receiver and transmitter is the ct axis translated to (x = 0, z = $z_0$). In Fig.3, at some time $t_1$ the transmitted (T) pulse is scattered off the point target. The receiver (R) detects only events on its own backward light cone (-t branch of the square root) at some time $t_0 + \tau$. The first return will occur at $t_2 = 2 z_0 /c$ after transmission (when $\tau \equiv 0$).

Conventional (real aperture nonimaging) radar in the following formulation can be considered as the limiting case of the preceding where $\tau=0$; that is, there is only one target at the nadir point and $z_0$ becomes the target range.

## Temporal Response of the Receiver

The receiver, in the standard time domain formulation of signal processing, produces an output voltage representing the convolution of the time-invariant serial receiver impulse response $\delta(t)$ with the incoming signal (6), here

$$\Gamma(\tau) = \int_{-\infty}^{\infty} \delta(t)\, \delta(t - \tau)\, dt$$

Generalizing to finite pulses, the temporal receiver response is:

$$\Gamma(\tau) = \Gamma(x(t), \tau) = \int_{-\infty}^{\infty} a_T(x, t)\ a_R(x, (t - \tau))\ dt \qquad\qquad \text{Eq.2}$$

where the spatial aspects of the response (curving of the world lines) are "hidden" in the $a_*$ functions. It should be noted that while $a_T$ and $a_R$ appear in symmetric roles in Eq.2, they are independent. The transmitter and receiver temporal characteristics are modelled as identical temporal rect( ) functions below.

A crucial point is that both $a_T$ and $a_R$ have different shapes (curvatures) at different values of their arguments, especially in the near field, thus changing the reception properties $\Gamma(\tau)$, without physical changes in the receiver and transmitter. As such, they are "hidden variables" in the integrand of $\Gamma(\tau)$. The integrand defines the shape and size of the spacetime region where the window worldlines $a_*$ overlap, or spacetime "channel" (7) which is the function $\zeta$;

$$\zeta(x,t,\tau) \equiv a_T(x, t) \, a_R(x, (t - \tau))$$

Through the spacetime channel, the transmitter can causally communicate with the receiver, though it will only do so in the presence of cooperative scatterers (at $x_1$ in Fig.3). Such scatterers deflect radiation from the outgoing (T) to the incoming (R) radiative worldlines both within the spacetime channel $a_T \, a_R$. Uncooperative scatterers would include, for example, specular areas that reflect radiation away from the receiver (e.g. deflect radiation into worldlines other than the parts of the R light cone that make up $a_R$).

## Radar Return Phenomena

The many properties of the returns from either a real-aperture, Doppler, or imaging synthetic aperture radar may be deduced by applying the preceding formulation to various observation scenarios.

## Spherical Pulse Worldlines: rect( ) Function

Far from the antenna, an emitted pulse is approximately spherical. Taking a typical satellite altitude ($z_0 = 800$ km), the outer and inner annulus radii $\sqrt{(ct)^2 - z_0^2}$ and $\sqrt{c^2(t - \tau_0)^2 - z_0^2}$ (respectively $a_T(t)$ and $a_T(t-\tau_0)$ in the above) are shown in Fig.4 as a function of time for an exaggerated pulse width of $\tau_0 = z_0/3c$. If the region between the lines is designated as having value unity, and the region outside is zero, this plot is the $a_T$ for a transmitter output window function that is rect($t/\tau_0$) on the time axis.

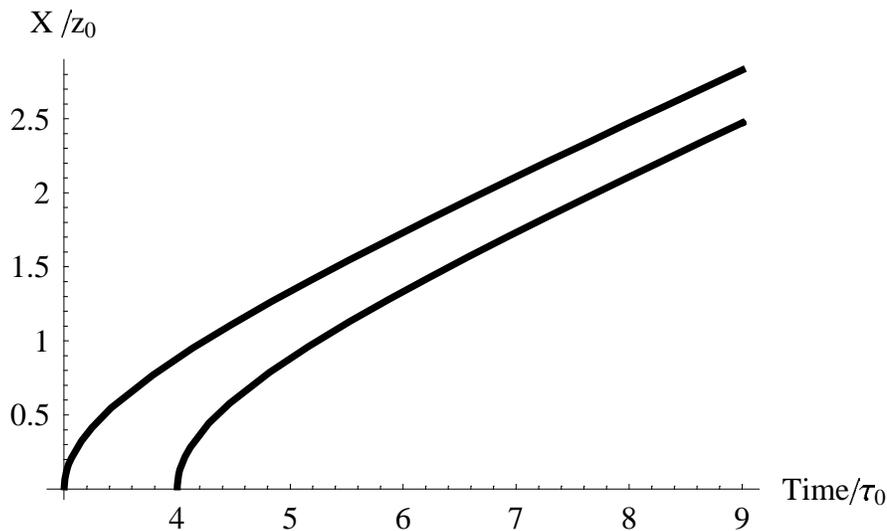

**Figure 4.** Spherical pulse radii $a_T(t)$ and $a_T(t-\tau_0)$ on the ground as a function of time for transmitter at 800 km altitude with exaggerated pulse length $z_0/3c$. The temporal (horizontal) separation between the lines is $\tau_0$ at all ranges x, approximately the inverse of the transmitter (or receiver) bandwidth.

In the region where x is comparable to $z_0$, there is significant curvature and change in pulse width; at far distances, width is approximately constant at $c\tau_0$. This radial width (difference between the two radii in Fig.4) of the illuminated region is presented in Fig.5; it increases from initial illumination at $t_0$ until the trailing edge of the pulse intersects the ground at $t_0 + \tau_0$, then decreases to its asymptotic value of $c\tau_0$.

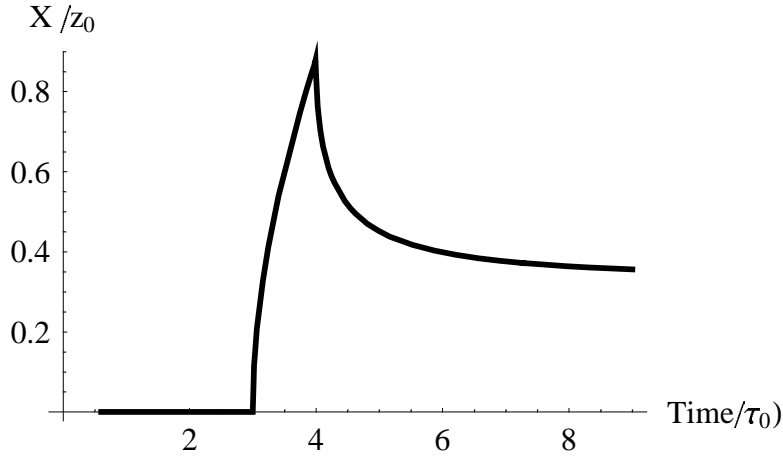

**Figure 5.** Radial pulse width as a function of time.

## Signal Strength

The geometric inverse-square attenuation of the pulse intensity is $1/\rho^2$ where $\rho = ct = \sqrt{x^2 + z_0^2}$. The power at the rezel is the inverse square of the distance from the transmitter (vertex of the T light cone). Then, for a spatially extended scatterer or "patch" of size greater than a wavelength, the intensity at the the receiver (vertex of the R light cone) is $1/(2\rho)^2$, and for a small (Rayleigh) scatterer it is $1/\rho^4$. In either case, the intensity is determined by the location of the rezel - the intersection of $a_T$ and $a_R$, which in turn depends on their functional shape (curvature)

## Monostatic Far Range: Radar and SAR Range Resolution

Eq.2 in one dimension (in the far range where the pulse width no longer varies) becomes a single variable form $a_*(x-ct)$. Then, at some arbitrary $x_0 \gg z_0$, Eq.2 becomes

$$\Gamma(\tau) \equiv \int_{-\infty}^{\infty} a_T(x_0 - ct)\, a_R(x_0 - c(t-\tau_0))\, dt$$

and the response of a scatterer that is visible (e.g. connects the T and R worldlines) to the receiver is then obtained by multiplying the integrand by some target reflectivity $\sigma(x_1, t)$, for simplicity $\sigma(x_1)$.

Choosing a unit rect( ) of temporal width $\tau_0$ and a stationary point target at some far range, $\sigma(x_1) = $ constant $\equiv \sigma$, the spacetime channel $\zeta(t,\tau)$ for a point target at $x_1$ in the far range $x_1$ becomes

$$\varsigma_t(t,\tau) \equiv \text{rect}(x_T(t)/c\tau_0)\ \text{rect}(x_R(t)/c\tau_0)\ \sigma(x_1)$$

where $x_*(t)$ is given by Eq.1 which in the far range becomes $x = \pm ct$ (temporal scaling is implicitly $t_0 = z_0/c$, which is independent of $\tau_0$).

The receiver output is the convolution $\Gamma$; the integration region is shown in Fig.6.

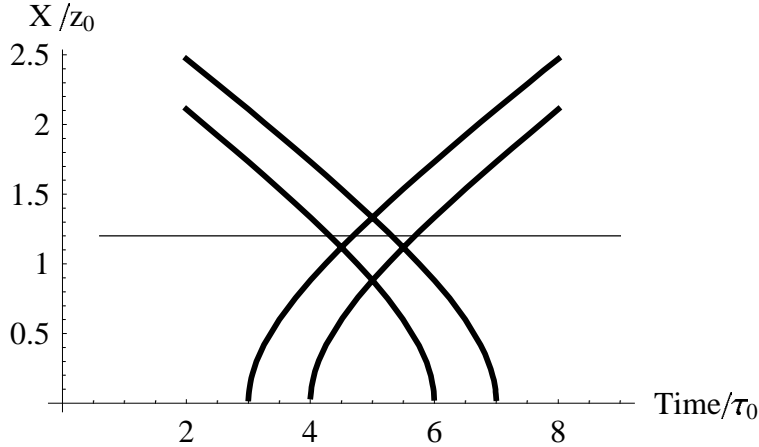

**Figure 6.** Components of the convolution $\Gamma$ that determine the receiver response of a point target for ERS ($z_0 = 800$ km) and an exaggerated pulse width $z_0/3c = \tau_0$. The transmitter rect( ) window worldlines $a_T$ are the two traces rising to the right; the receiver rect( ) window worldlines $a_R$ fall to the right. The offset variable $\tau$ is measured along the time axis as the distance between the outermost two traces. The diamond shaped area is the spacetime channel $\varsigma$; the horizontal line is the worldline of a target at some distance $x_1$.

The integrand is only nonzero in the spacetime channel $\varsigma$; the full integrand including target $\varsigma_1$ is therefore nonzero only where the target worldline intersects $\varsigma$. $\Gamma(\tau)$ is thus the length of target wordline segment inside the spacetime channel. From the diagram, in the far range,

$$\Gamma(\tau) = \sigma\ \text{tri}((c\tau)/2c\tau_0)$$

where the spacelike (x) extent of the channel must be $2c\tau_0$. The full width half maximum of this peak is $c\tau_0$, the usual far range resolution of a pulsed radar.

## Monostatic Near Field: Radar and SAR

Eq.2 is now used to examine the near and intermediate field imaging properties. The outgoing $a_T$ of the transmitter pulse and the incoming $a_R$ of the window of the receiver (worldlines) were shown in different contexts in Figs. 3-6. Figs.7, 8, and 9 show them again to indicate graphically the integration area (spacetime channel) of $\Gamma(\tau)$ for three different $\tau$'s representing the near, intermediate and far range. As previously, there is a single spatial dimension x and a unit rect( ) function pulse, and an exaggerated pulse length $\tau_0 = z_0/3c$ for $z_0 = 800$ km; the light area in the plots is the spacetime channel $\varsigma$.

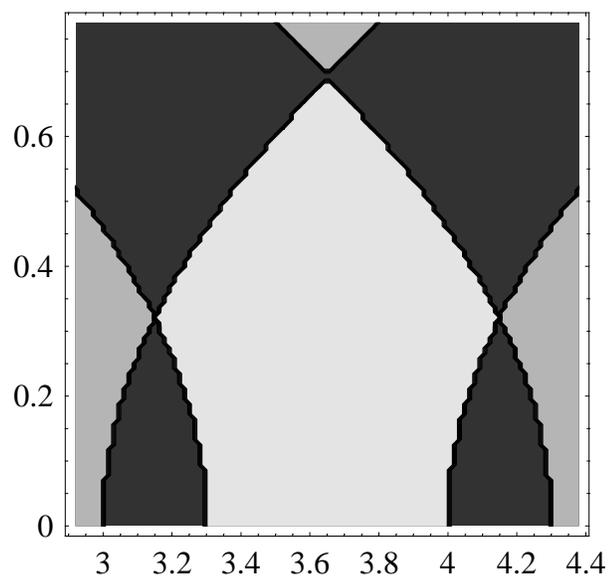

**Figure 7.**

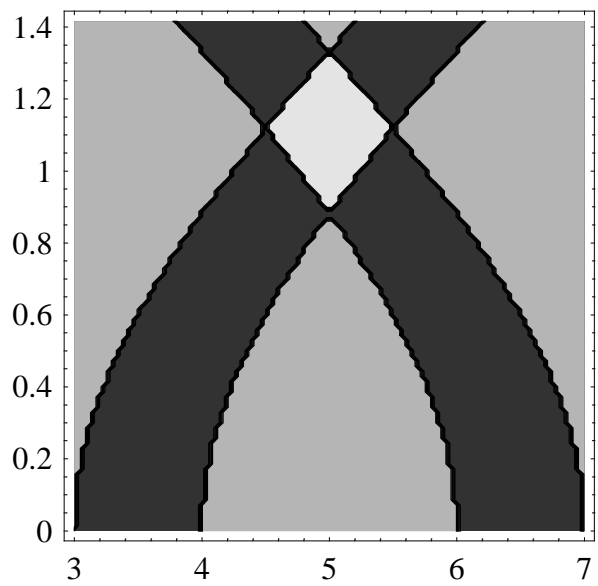

**Figure 8.**

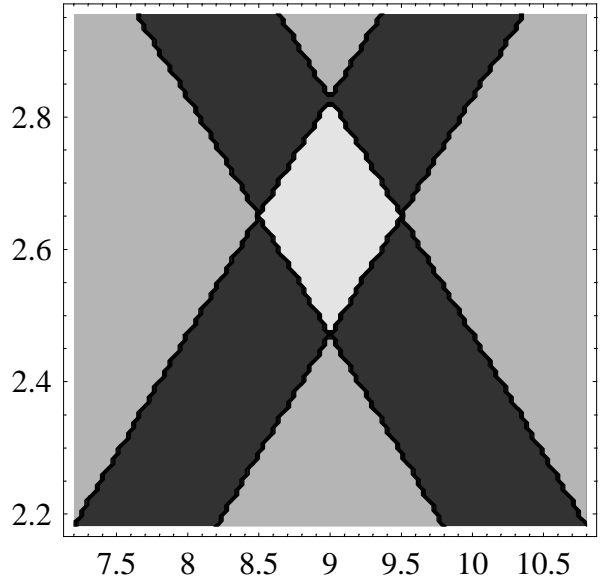

**Figure 9.**
**Figures 7, 8, 9.** Horizontal axes are in units of $\tau_0$; vertical axes in units of $z_0$. Areal overlap of the integration of Eq.2 as a function of the offset, $\tau$, scaled to show the spacetime channel. Fig.7 is near the origin (x=0) where $\tau$=0, Fig.8 in the intermediate region where $\tau$ is a low multiple of $t_0$ and Fig.9 is a closeup of the far range, $\tau \gg t_0$. Each plot contains a finite-pulse $a_T$ and $a_R$; the spacetime channel is the light area is where both functions are nonzero; the dark areas are where one or the other function is nonzero, but not both; the remaining 'background' areas are where both functions are zero. Only the light area contributes to the integral. Fig.9, which is itself a limited region of Fig.6 without a target and represents the spacetime region in which most radars are designed to operate.

As in Fig.6, the x dimension of the spacetime channel determines the resolution element or "rezel" size, which is the spacelike dimension of the light region, the spacetime channel $\varsigma$. This picture extends in a straightforward way to the case of a long pulse coded waveform radar. The total area of $\varsigma$ determines the total possible scattering strength. If, for example, the scene consists of uniform isotropic (connect any incoming radiation to all outgoing worldlines) scatterers, a large spacetime channel (for example at the nadir shown in Fig.7) will produce a large receiver output $\Gamma$. This variation of the signal strength depends only on the spacetime channel properties determined solely by the height $z_0$ and the T and R approximate bandwidth $1/\tau_0$.

Fig.10 shows the results of numerical integration of Eq.2, which clearly shows the variation of $\Gamma$ over a flat scene.

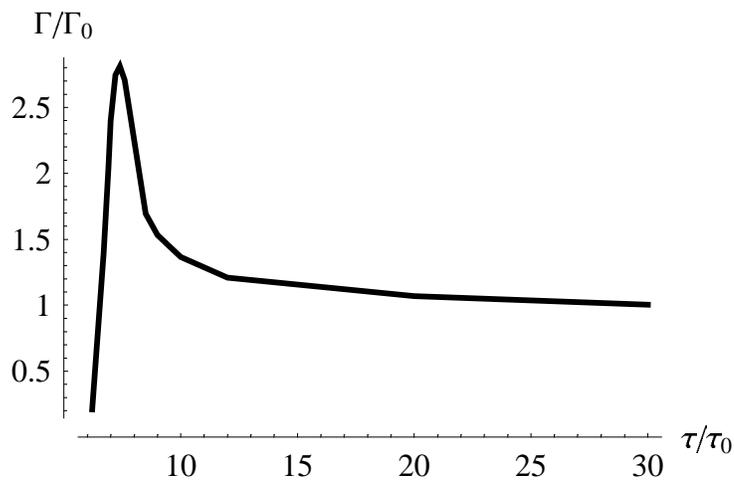

**Figure 10.** Numerical size of spacetime channel (ambiguity) region for uniform scatterers for a monostatic SAR at altitude 800 km with exaggerated pulsewidth $\tau_0 = z_0/3c$ as a function of separation $\tau$.

In the far range the rezel size is approximately constant, as indicated by the asymptotic behavior of $\Gamma(\tau)$, which is also the usual operating region for a radar as in Fig.9.

It is clear that what is usually called "resolution" (the rezel size) can be interpreted as the smallest unresolvable ambiguity, although the term is usually used to mean only the far range spacelike value $c\tau_0$. Further, the spacetime channel and the related resolution are properties of both the transmitter *and receiver*, a notion that becomes more obvious in bistatic configurations (below).

Fig.11 indicates how the effect of topography on $\Gamma(\tau)$ can be included by applying a rotation to the coordinate system. A scene element is taken to be at angle $\gamma$.

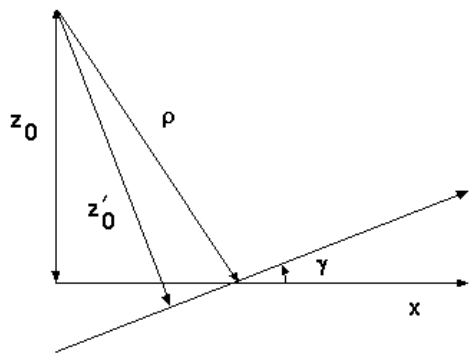

**Figure 11.** A rotation $\gamma$ can be viewed as defining a new radar height $z_0 \to z_0'$, which changes the imaging properties.

In the one dimensional case, a part of the scene tipped at some oblique angle $\gamma$ toward the radar defines a new height above the nadir, say $z_0'$ as indicated in Fig.11, and a corresponding new $t_0' = z_0'/c$. Then $\Gamma$ for that scene area (x) is given by the recalculated (with $z_0'$) curve in Fig.10 ; stated differently, the worldlines $a_s$, which depend on $z_0$ change with $\gamma$, changing $\zeta$, $\Gamma$ and the associated rezel size for that scene area. In the case where the scene slope is approximately (within the pulse width $c\tau_0$) perpendicular to the radar look direction, $\Gamma$ is recalculated for a new subradar point (e.g. $z_0'$), producing a large rezel (Fig.7). This situation is usually termed "layover", a form of ambiguity characterized by large rezels and correspondingly bright echoes. Topography profoundly influences layover; also clear from the foregoing is how layover ambiguity, like resolution, is influenced by both the transmitter *and receiver,* though it is usually thought of as a property of the transmitter only. This large rezel that occurs perpendicular to the look direction is a special case of the general fact that spatial resolution as a function of look angle $\gamma$ is not constant (Figs.7 and 10 make this clear). Resolution as a function of $\gamma$ is a nonlinear rescaling of the horizontal axis to reflect $\tau_0' = z_0'/c$ where $z_0' = z_0 \cos \gamma$. Thus $\Gamma$ as a function of $\gamma$ looks approximately like an inverted cusp constructed by ignoring the increasing part of Fig.10 and reflecting the decreasing part around the vertical line through the peak where $\gamma = \pi/2$. This same scenario over flat scenes ($\gamma$=0 but $\rho$ changes) produces a changing rezel size as x increases, which is called slant range distortion.

# Bistatic Far Range

Eq.2 can be applied to bistatic imaging by allowing two different x locations (two different worldlines) for the transmitter and the receiver, calculating the corresponding worldlines $a_T(x_1(t))$ and $a_R(x_2(t-\tau))$ and performing the integration (scatterers are hereafter considered constant and global in the scene).

A spacecraft at $z_{0T} = 800$ km is again used for $a_T(x_1(t))$; the receiver is assumed to be at $z_{0R} = 50$ km , which defines $a_R(x_2(t-\tau))$. The transmitter and receiver are assumed, as above, to have a common $\simeq 1/\tau_0$ bandwidth, and are also assumed to be at the same x (spacecraft directly over aircraft). For this case, Fig.12 shows the functions $a_T(x_1(t))$ and $a_R(x_2(t-\tau))$ in analogy to Fig.3.

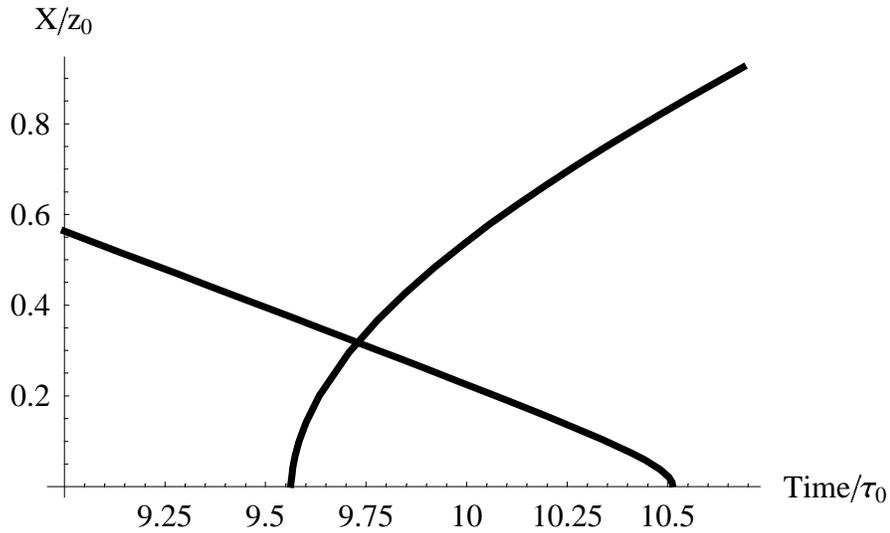

**Figure 12.** Bistatic outgoing pulse worldline $a_T$ for transmitter on a spacecraft and time reversed or "incoming" window $a_R$ for a receiver at a lower altitude for a very short pulse. The two curvatures are different because the receiver height $z_0$ is 50 km versus 800 km for the transmitter (Eq.1).

For finite pulses, the spacetime channel or region where the receiver can communicate causally with the transmitter, is thus somewhat different from the monostatic case shown Fig.8; but in the very far range, both worldlines must have the same slope, and the intersection area for finite pulses approximates that of Fig.9, so the resolution is the same as the monostatic case.

Fig.13 depicts the illumination of the scene of Eq.2 for another bistatic configuration. X is horizontal and y is vertical in the page, with the plot scaled differently in x and y. The two windows $a_T$ and $a_R$ are shown in the two scene dimensions (light cone intersects (x,y), not (x,t)) in the intermediate region $z_{0R} \lesssim r \ll z_{0T}$ and with the transmitter distant from the receiver in both z and x.

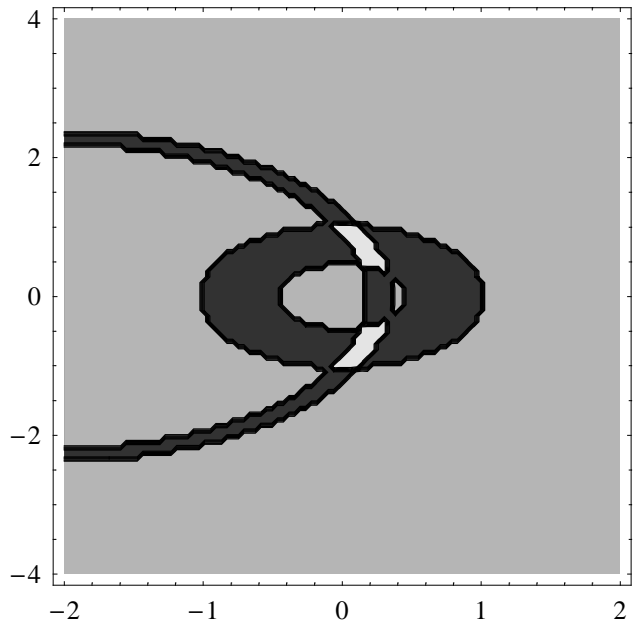

**Figure 13.** Image plane (x,y), distances in meters. (x and y are scaled to $z_0$). Finite pulse worldlines $a_T$ and $a_R$ intersection with the plane (x,y) for a bistatic configuration (either ring can be T or R, with the other then R or T). The bistatic subradar points differ, accounting for the spatial offset of the rings. The relative sizes of the rings is determined by the offset $\tau$ (see Fig.2).

The bright (both functions nonzero) areas, for unit pulses indicate that $\Gamma$ has two branches. Thus, a bistatic range rezel may not be simply connected even over a flat scene, it here has two branches, a form of ambiguity. Similary, a monostatic radar may have multiply connected rezels (ambiguities) over nonflat targets.

# Bistatic Ambiguity

An anomalous case of bistatic imaging occurs when the area between two separated antennas is imaged. If both the transmitter and receiver are at some altitude $z_0$ and separated by some x, the functions $a_T(x(t))$ and $a_R(x(t - \tau))$ of Eq.2 are almost identical, but have a spacelike (x) separation as depicted in Fig.14.

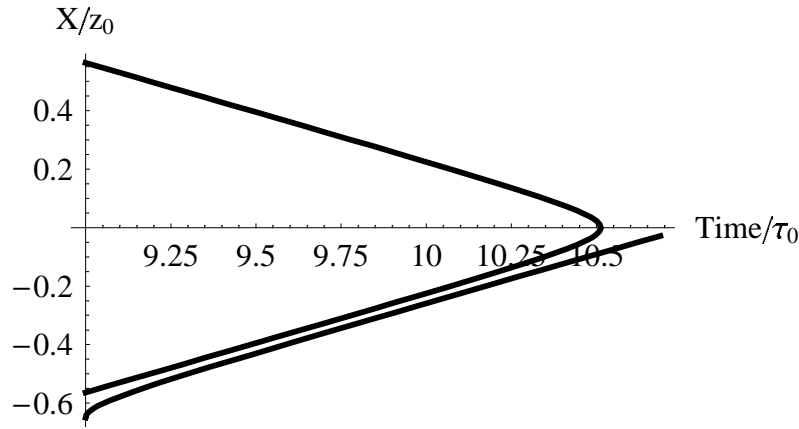

**Figure 14.** Bistatic worldlines $a_T$ and $a_R$ calculated from (1) for the case of imaging of the area between a separated transmitter and receiver. The lowest line is $a_T$, here offset in the -x direction by some amount that reflects the distance from the transmitter's nadir to the receiver's nadir (lower square root branch is suppressed); the upper two connected lines are the two square root branches of $a_R$ (both branches of the intersection of the incoming $a_R$ light cone with the (x,t) plane). A slight relative shift (change in $\tau$) from what is in the image would create a huge spacetime channel (the two lower lines would coincide over a large area).

Here an interesting thing happens; the *lower* (-x) branch of the curve $a_R$ intersects the area of the scene lying between two antennas, and can be almost identical to $a_T$. Even in the far range, where the $a_*$ functions are simple, $\Gamma(\tau)$ will have one very narrow (width $\approx c\tau_0$) peak of very large magnitude. Interpreting Fig.14, this is because the receiving worldine $a_R$ (t-$\tau$) overlaps substantially with the illuminating worldine $a_T$(t); physically the outgoing radiation window where $a_T$ on the ground moves in almost the same way as the incoming window $a_R$, producing coincident tracks similar to Fig.14. Thus the spacetime channel becomes very large. As a rough calculation, if the antennas are separated by $x_0$, there are approximately $x_0/c\tau_0$ rezels, which, multiplied by the far range monostatic value of $\Gamma(\tau)$ in Fig.10, yields the approximate value of $\Gamma(\tau)$ for this bistatic configuration, a number much larger than the equivalent peak monostatic value. The large spatial extent of the integrand in $\Gamma(\tau)$ indicates poor resolution, that is, returns from large ranges of x only for a very limited range of $\tau$, the convolution offset..

Accordingly, the rezels between the antennas are huge, resulting in a very bright, short received signal and poor resolution. A more conventional explanation is shown in Fig.15, which symbolically depicts the radar scene and two ellipsoidal ambiguity surfaces in the (x,z) plane.

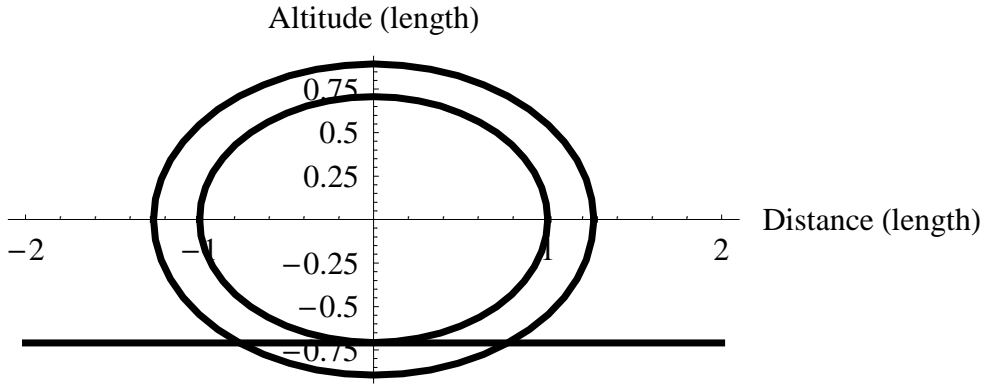

**Figure 15.** Ellipsoidal ambiguity surfaces for bistatic radar in the (x,z) plane. The T and R antennas are the foci of the ellipses; the lower horizontal line is the x dimension of the scene (the ground).

The ellipses are defined by the two antennas located at the foci; the scene is represented by the lower horizontal line. The time required for a signal originating at one antenna to reach any point on either ellipse and then be received at the other antenna is independent of the location on the ellipse; thus the two points where the outer ellipse intersects the ground are ambiguous. If the outer ellipse represents a propagation time $\tau_0$ later than the inner ellipse, any parts of the scene between the two ellipses are in the same rezel and are hence ambiguous. This is the physical reason why it is difficult or impossible to image the area between two antennas.

## Pulse Limited vs. Beam Limited Systems

In general, "beam limited" and "pulse limited" radiators appear to operate quite differently. The one is radiating for times much longer than the dimensions (e.g. $z_0$) associated with the observation. The latter, like SAR in the range dimension, defines a spatial scale by the pulse width that is, loosely speaking, much smaller than the illuminated area. The foregoing formulation can be applied to pulse limited or beam limited systems merely by changing pulse length $\tau_0$ when constructing the functions $a_T$ and $a_R$. Short $\tau_0$'s that represent small simultaneous scene coverage give ambiguities and resolutions that are pulse limited; large ones that produce large coverages yield beam limited characteristics. There are intermediate regions with hybrid characteristics for both the transmitter and the receiver.

## **Coherent Phemonena and Interferometry**

The foregoing analysis has been, implicitly, scaled to the length $z_0$. Now it is applied to wavelength-dependent phenomena where there is the independent length $\lambda$ and corresponding period $c/\lambda = \nu$.

### Phase and Speckle

Phase is time, usually measured modulo $2\pi$ at some carrier angular frequency $\omega_0$, and can be viewed as the scale of the t axis in Eq.2. In Fig.3, the one dimensional far range depiction of Eq.1, phase can be identified as the world line time (t axis) of the receiver measured in units of the carrier frequency. In the colloquy of radar, it is implicit that receiver "phase" refers to the far range, even though this is only a limiting case. The measured phase value associated with a rezel depends on the occurrence of the sampling event at the receiver - that is, the loca-tion of the receiver window on the time axis. The chosen global synchronization of the sampling, usually designated $\phi_0$, refers to the length of the receiver world line for far range light areas (Fig.9) when the returning pulse is temporally centered on the receiver; this is shown symbolically in Fig.16 (a finite antenna or phased array could be accounted for by a spacelike addition of worldlines over the antenna surface and a timelike addition over postreception phase shifts).

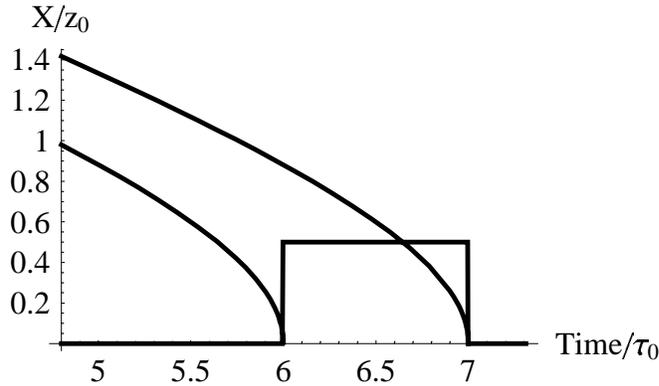

**Figure 16.** Schematic representation of reception; two incoming radiation worldlines $a_R$ framing a flat-scene rezel (e.g. separated temporally by the inverse of the transmitter bandwidth) are shown with the model of the receiver temporal aperture. The receiver integrates across the interval defined by the same bandwidth as the pulse; the receiver phase associated with the indicated rezel is definitionally measured when the receiver is synchronized with and integrating the indicated interval.

The incoming radiation is measured against the local oscillator in the receiver. Each resulting phasor measurement is then added to (more correctly, integrated with) the other phasors received in the same bandwidth-limited window. That coherent sum (integral) yields the output phase of the receiver. In mixed terminology, it is the cross-correlation of the receiver response with the worldlines $a_R$ (x=0)) of the incoming radiation.

In reality, a rezel contains many scatterers which result in a random distribution of worldlines and phases. In the case that the rezel can be modeled as a collection of differently oriented patches with size greater than a wavelength (e.g. non-Rayleigh scatterers), the different orientations each correspond to a different $z_0$' as shown in Fig.11, which in turn implies different worldlines $a_*$ and for each patch. The result is that the spacetime channel in Fig.9 has a noiselike phase that is a superposition of the multiple (varying curvatures) worldlines, measured and added coherently in the timelike (time serial) receiver. The randomness of the incoming worldlines then produces a random output phase, usually referred to as speckle, although this terminology is rather imprecise since even a single patch, being spatially finite, has slightly varying worldlines associated with it.

While a sampling event will always produce a phase, if the measured rezel is larger (Fig.7) than the far range rezel (Fig.9), the measurement represents in a sense an ill-posed interrogation of the receiver; the assumed simple conditions of Fig.9 *around which phase is defined* have been violated. Similarly, in the case of a "multipulse" ambiguity, $a_R$ would intersect two or more distinct nonzero $a_T$ regions associated with multiple radar pulse , again resulting in a phase that has two or more contributing sources; that is, the receiver would output some phase, but it is meaningless in the sense that it does not fit the colloquial definition.

 In spacetime terms, phase is the difference between the time on the world line of the receiver ($\tau$) and the time measured by a clock (e.g. radiation phase) travelling on the world line of the moving pulse. This is the same time difference as that in the twin paradox of special relativity; in both cases a local clock is compared to a moving clock which goes to some distant point and returns. The twin paradox states that because of time dilation, a moving (twin) clock must return showing an earlier time than the local clock, which implies that the frame of reference of one clock is somehow different ("preferred"), which is impossible. The paradox is resolved by noting that the moving clock must experience an acceleration to return to the other clock, so it indeed has a preferred (noninertial) frame of reference. Scattering, which changes the worldlines of the radiation from $a_T$ to $a_R$ plays the role of the "acceleration" that makes one twin noninertial; geometric speckle is the radar expression of the twin paradox .

## SAR Azimuth Ambiguity

In the simplest case of azimuth (sampling) ambiguity, a target also at height $z_0$ has a small x velocity, and a world line that is slightly pitched compared to the world line of the receiver. The receiver only measures time from the nearest $2\pi$ phase multiple; Nyquist subsampling may disguise a phase change of more than $2\pi$. As depicted in Fig.17, this results in the receiver "seeing" the target as if it were on the wrong world line.

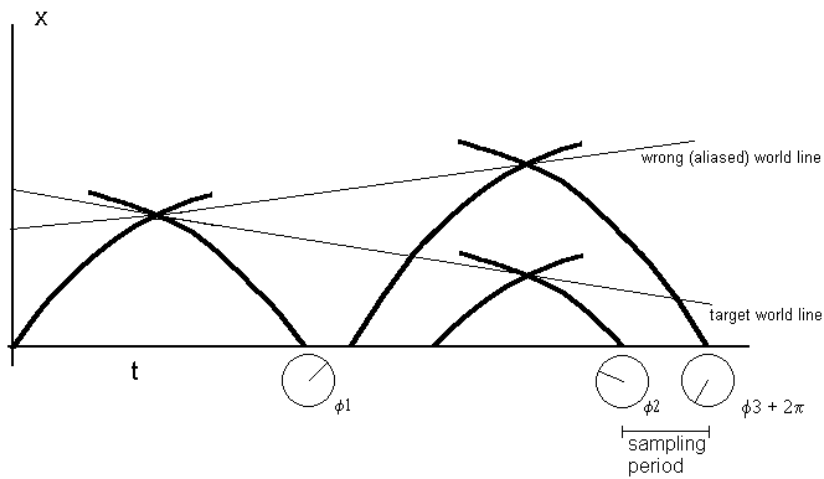

**Figure 17.** Worldlines associated with azimuth aliasing by receiver. The circles indicate the phase measurements $\phi_1$, $\phi_2$, $\phi_3$ made at discrete, evenly spaced sampling times. The third measurement is assumed to include a subsampling error of $2\pi$, causing the receiver to produce a false target in the image, which can be thought of as a ghost wordline in spacetime.

In Fig.17, the rightmost incoming $a_R$ is assumed to be sampled by a receiver operating at half the Nyquist rate, causing a phase error of $2\pi$. The aliased worldline is precisely what appears in a radar image - a false target at a location determined by the sampling frequency. In radar terminology it is "on the wrong ambiguity" (3). Subsampling at some arbitrary frequency may produce multiple diffractive ghost worldlines, each corresponding to some resonance of the sampling frequency and spectral component of the incoming signal.

# Interferometric Configurations:
# Repeat Orbit, Cross Track, Along Track, and Deformation

Interferometric radar ( for example (8)), is the limiting case of bistatic radar where the two antennas, T and R, are almost colocated, separated by a small distance $B \ll z_0$ referred to as the baseline. Alternatively, interferometry can be performed with repeated observations from an orbital radar, where the spatial difference between the orbits is the baseline.

The integration of Eq.2 for interferometry is represented in analogy to Fig 3 by Fig.18, where the two different curve pairs $c_{T1}$, $c_{R1}$ and $c_{T2}$, $c_{R2}$ represent the two observations ($c_{**}$ refers to the curve pair $a_{**}(t)$ and $a_{**}(t-\tau)$).

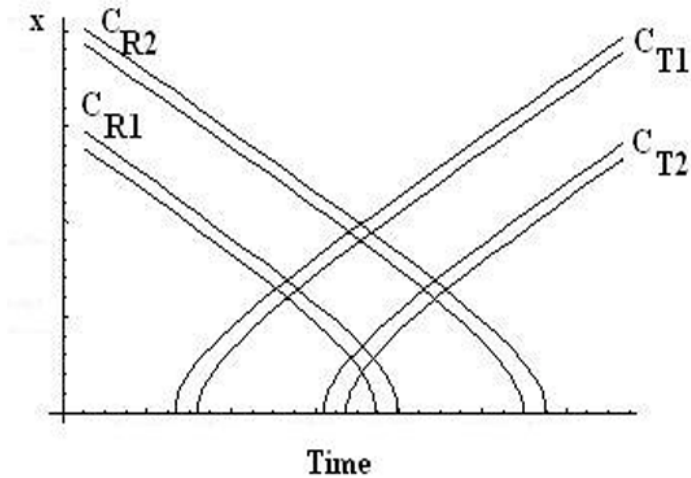

**Figure 18.** Worldline pairs for interferometric radar. Each pair $c_{T*}$ is composed of $a_T(t)$ and $a_T(t-\tau)$; $c_{R*}$ is composed of $a_R(t)$ and $a_R(t-\tau)$.

Parallel coflying antennas ("crosstrack" interferometry) are represented in this picture by removing one of the upward sloping curve pairs; that is, there are one transmission ($c_{T1}$ and not $c_{T2}$), then two receptions ($c_{R1}$ and $c_{R2}$) spatially separated by B.

Along track interferometry (9) can be represented by Fig.18 where the curvatures of $c_{*1}$ and $c_{*2}$ are the same. The platform velocity (across x, or in the y direction) positions the antennas at the same x separated by the timelike interval $\tau_{\text{across}}$. In the usual application (ocean current measurement), the scene has a rezel moving at uniform speed v parallel to x. This translates into a rezel offset of $v\tau_{\text{across}}$, resulting in observations at two different times, at two different locations or curvatures of $c_{1x}$, producing a phase shift from which v can be inferred.

Deformation interferometry (10) can be visualized in a straightforward way with a third curve pair $c_{3*}$ and is discussed below.

# Interferometric Phase, Baselines, Height and Finite Antenna

The curvatures of the worldline pairs $c_{*1}$ and $c_{*2}$ differ slightly in interferometry. This difference comes from the spacelike (cross track or vertical) separation of the antennas separation, and is the origin of interferometric phase, which in Fig.18 is the *change* in the separation $\tau$ between the c curves measured along the receiver world line(s).

In the simplest picture, two antennas are separated in height z by an amount B, the interferometer baseline. Picking the "1" pair as the reference, a rezel has height that can be interpreted as a different $z_0$, a different set of curvatures in the $c_{**}$ functions and thus a new (interferometric) phase corresponding to that rezel. Since the baseline B $\ll z_0$, a change in the length of B can be approximated as a rescaling of the vertical (spacelike) axis in Fig.18 by the factor $1+\Delta B/z_0$ for the $c_{*2}$ pair. The resulting far range linear divergence between $c_{R1}$ and the rescaled $c_{R2}$ as x increases is precisely what was seen in (10) which had images of the very flat Imperial Valley of California, where there were no fringes due to topography (e.g. z0 is constant).
Any phase effects resulting from a finite antenna size can be treated with the same approach; world lines can be constructed from a baseline that now represents the spatial distance between antenna elements. This is in the x direction for the one-dimensional case under consideration, but can be generalized to three spatial dimensions and may even include timelike shifts to account for propagation delays within the antenna.

Finally, a height shift (deformation) of a particular rezel between observations can also be represented as a change in $z_0$, $\Delta z_0$ resulting in an *additional* curvature change of, say, $c_{*3}$ only (that is, the deformation occurred in the 2→3 time period). The required processing step in deformation SAR of phase rescaling (10) approximates the removal of the different curvatures by changing the far range slopes of two curves $c_{*1}$ and $c_{*2}$ to match $c_{*3}$. A globally constant phase and a phase shift from the *residual* curvature changes in $c_{*3}$ caused by the deformation $\Delta z_0$ expresses itself as a phase change (10) in a "double difference interferogram".

# Interferometric Decorrelation

If the two interferometric phase measurements speckle in an uncorrelated manner for any reason for a given rezel, interferometric phase differencing breaks down. If the source of the speckle is, for example, layover the worldlines $a_T$ and $a_R$ become of indeterminate (scene-dependent) curvature and location (Fig.11), and the either phase can change rapidly with baseline; the interferometric phase becomes indeterminate, a phenomenon usually called "decorrelation". However, if the baseline is zero, the phases measured by the receiver are still deterministic in the sense that they can be reproduced. Interferometric coherence necessitates that two spacetime channels (usually shaped like in Fig.9) must overlap almost completely; sufficiency occurs when the sampling events (phases) on the receiver worldline represent almost the same sequencing of scene ($\gamma$) dependent worldlines which is when there is *both* overlap and an undisturbed scene. Small changes in the spacetime channel, like those produced by small baselines, will produce an almost non-noisy phase *difference* at the receiver, which will then generate high SNR interferometric fringes.

Large baselines cause decorrelation because they result in spacetime channels so with so much parallax (changes in $z_0$ and $\gamma$ in Fig.11 for the scatterers within a pixel) that the sequencing of worldlines seen by the receiver changes to the point of producing uncorrelated output phases. In engineering terminology, the random worldlines originating in the spacetime channel overlap (common area) is called common mode noise, which can be eliminated between two observations if the receiver sum (i.e. worldline sequencing on the time axis) does not change profoundly. The interfometric SNR can only be unity when there is exact spatial alignment of observations.

Other sources of decorrelation - multiple pulse ambiguity, topography, and topographic interference fringes in double difference interferometry can all be described in the same way, as properties of the spacetime channel, which gives a general picture of when it is possible to eliminate common mode noise and derive a high SNR interferometric phase.

Surface disruption on scales less than one rezel, can be thought of in this framework as a redistribution of the 'patches' (asserted above to be the source of non-Rayleigh speckle) and implicitly, the undecipherable scrambling of scatterer worldlines.

### Nyquist Rate for Interferometric Fringes

Steep terrain, or equivalently, a large baseline, makes the fringes representing the interferometric phase difference occur more rapidly than the (spatial) Nyquist rate (one fringe per two rezels) and decorrelation results from the consequent subsampling and aliasing. In the picture of Fig.18, the curvature of $c_{*1}$ differs significantly from that of $c_{*2}$ because of a large rezel-to-rezel change in $z_0$, the interferometric phase measured between the two intersections can change by more than $\pi$, just like azimuth ambiguities. This is comonly observed in SAR interferograms.

### Doppler Shift: Moving Targets and Platform Motion

Radar Doppler shifts are usually associated with a moving target that reflects radiation. In the picture presented by Fig.9, the outgoing $c_{T*}$ will have a longer intersection with a target worldline that is pitched with respect to the t axis; similar comments apply to $c_{R*}$. Accordingly, the receiver phase changes . Dimensionally, for small v, the first order slope change of the light areas must be $(1 \pm v/c)$, resulting in the familiar Doppler multiplier of the carrier frequency. Similarcomments apply if the platform is moving and thus has a pitched worldline.

### Atmospheric Effects

In the simplest case an atmospheric disturbance consists of a medium, such as water vapor, uniformly covering the scene and propagation paths. In the spacetime methodology, this can be represented by a stretching of the time axis, since the speed of light c becomes c/n in medium of index of refraction n. More complex scenarios of spatially inhomogeneous gases can be dealt with in an obvious way, where the world lines of radiation follow the local spacetime geodesic.

## Geometric Transformations of the Spacetime Channel

The various experimental configurations described above correspond to simple geometric transformations that determine the morphology and orientation of the spacetime channel $\zeta$. Resolution or ambiguity is determined by the translation $\tau$ (Fig.10) or the rotation $\gamma$ that defines layover. Atmospheric effects are a time dilation; the two bistatic configurations are spatial translations of the T and R light cones. Similarly, coflying interferometry is a rotation of the white area of Figure 9 by the angle $\gamma$ that defines some baseline; repeat pass interferometry is a space translation as well; along track interferometry is a time translation; deformation interferometry is two rotations and two time translations. Doppler shifts can be viewed either as a translation or a dilation.

## Fundamental Physics

### Hidden Variables in Relativistic Spacetime

A circle projected onto a plane can appear as a circle, an oval, or a line, depending on the orientation in three dimensions. The orientation of the circle is a "hidden variable", which can be inferred in the two-dimensional plane but not directly observed. Even simple hidden objects can produce complex behaviors in lower dimen-sional spaces - for example the planar projection of the line segments that constitute a rotating cube.

In the foregoing discussions of radar, the worldlines of the transmitter, receiver, pulses and scatterers were shown to account simply for many seemingly unrelated radar phenomena - ambiguity, layover, resolution, layover, speckle, obliquity, signal strength, beam and pulse limited configurations, Doppler shifts, finite antenna size, and atmospheric delays - even in very different experimental configurations. A simple examination of changes in the worldlines $a_T$ and $a_R$ in interferometric configurations was used to explain more known but seemingly unrelated properties - interferometric resolution, decorrelation, surface motion, and surface deformation. *Many of the seemingly unrelated complex phenomena of radar returns are just the projection of simple higher dimensional hidden variables - the intersections of the worldlines $a_T$ and $a_R$ - onto the one-dimensional (time serial) receiver.*

At a more fundamental level, the curvature of the worldlines $a_*$ is what determines many of the complex behaviors of radar. This is also what happens in general relativity; radiative worldlines that deviate from the light cone are identified with many of the most complex and interesting cosmic phenomena, including, for example, black holes.

The curved worldlines $a_*$ themselves result from a broken symmetry - a radar pulse with spherical symmetry intersecting a target, resulting in a lower circular symmetry. This broken symmetry is the fundamental cause of the complex phenomena observed in radar signals.

## Summary and Conclusion

A formalism was developed that describes the properties of radiative imaging systems. Many complex phenomena of radar - resolution, ambiguity, speckle, obliquity, beam and pulse limited configurations, interferometric resolution and decorrelation, and Doppler shifts properties were shown to originate in the curvature of the world line of the transmitter and receiver pulses, or more abstractly, higher dimensional hidden variables in Minkowski space. This explanation, while abstract, is actually quite useful for understanding radar returns and optimizing experiments, since it provides both a unified and simple view of disparate phenomena.

## Acknowledgements

We wish to acknowledge the support and patience of Paul Rosen, and the interest and encouragement of Giorgio Franceschetti.
The research described in this paper was carried out at the Jet Propulsion Laboratory, California Institute of Technology, under a contract with the National Aeronautics and Space Administration.